\documentclass[final,5p,times,twocolumn]{elsarticle}

\usepackage{rotating,color,subfigure,amssymb}
\usepackage{alphalph}
\journal{Physics Letters B}

\begin{document}

\begin{frontmatter}

% Title, authors and addresses

% use the tnoteref command within \title for footnotes;
% use the tnotetext command for theassociated footnote;
% use the fnref command within \author or \address for footnotes;
% use the fntext command for theassociated footnote;
% use the corref command within \author for corresponding author footnotes;
% use the cortext command for theassociated footnote;
% use the ead command for the email address,
% and the form \ead[url] for the home page:
% \title{Title\tnoteref{label1}}
% \tnotetext[label1]{}
% \author{Name\corref{cor1}\fnref{label2}}
% \ead{email address}
% \ead[url]{home page}
% \fntext[label2]{}
% \cortext[cor1]{}
% \address{Address\fnref{label3}}
% \fntext[label3]{}

\title{Measurement of the $\eta\to 3\pi^0$ Dalitz Plot Distribution
with the WASA Detector at COSY }
\author{The WASA-at-COSY Collaboration}

\author[Erl]{C.~Adolph}
\author[IKPJ,JCHP]{M.~Angelstein}
\author[PITue]{M.~Bashkanov}
\author[IKPJ,JCHP]{U.~Bechstedt}
\author[FBGat]{S.~Belostotski}
\author[ASWarsH]{M.~Ber{\l}owski}
\author[IITB]{H.~Bhatt}
\author[HISKP]{J.~Bisplinghoff}
\author[Budk]{A.~Bondar}
% ...
\cortext[aa]{Corresponding author} \ead{p.vlasov@fz-juelich.de}
\author[IKPJ,JCHP,Katow]{W.~Weglorz}
\author[MS]{A.~Winnem\"oller}
\author[IKPJ,JCHP]{A.~Wirzba}
\author[IKPJ,JCHP]{M.~Wolke}
\author[IPJ]{A.~Wro{\'n}ska}
\author[ZELJ,JCHP]{P.~W\"ustner}
\author[IMPCAS]{H.~Xu}
\author[KEK]{A.~Yamamoto}
\author[KEK]{H.~Yamaoka}
\author[IKPJ,JCHP,IMPCAS]{X.~Yuan}
\author[IKPJ,JCHP,NuJINR]{L.~Yurev}
\author[ASLodz]{J.~Zabierowski}
\author[IKPJ,JCHP,IMPCAS]{C.~Zheng}
\author[IKPJ,JCHP,IPJ]{M.J.~Zieli{\'n}ski}
\author[Katow]{W.~Zipper}
\author[IKPUU]{J.~Z{\l}oma{\'n}czuk}
\author[ZELJ,JCHP]{K.~Zwoll}
\author[ASSwi]{I.~Zychor}

\address[Erl]{Physikalisches Institut,
    Universit\"at Erlangen--N\"urnberg,
    Erwin--Rommel-Str.~1, 91058 Erlangen, Germany
    }
\address[IKPJ]{Institut f\"ur Kernphysik,
    Forschungszentrum J\"ulich,
    52425 J\"ulich, Germany
    }
\address[JCHP]{J\"ulich Center for Hadron Physics,
    Forschungszentrum J\"ulich,
    52425 J\"ulich, Germany
    }
\address[PITue]{Physikalisches Institut der Universit\"at T\"ubingen,
    Auf der Morgenstelle 14, 72076 T\"ubingen, Germany
    }
\address[FBGat]{The Few Body System Laboratory,
    High Energy Physics Division,
    St.~Petersburg Nuclear Physics Institute,
    Orlova Rosha 2, 188300 Gatchina, Russia
    }
\address[ASWarsH]{High Energy Physics Department,
    The Andrzej Soltan Institute for Nuclear Studies,
    ul.\ Hoza 69, 00-681, Warsaw, Poland
    }
\address[IITB]{Department of Physics,
    Indian Institute of Technology Bombay,
    Powai, Mumbai, 400076 Maharashtra, India
    }
\address[HISKP]{Helmholtz--Institut f\"ur Strahlen-- und Kernphysik,
     Rheinische Friedrich--Wilhelms--Universit\"at Bonn,
     Nu{\ss}allee 14--16,
     53115 Bonn, Germany
     }
\address[Budk]{Budker Institute of Nuclear Physics,
    akademika Lavrentieva
    prospect 11, 630090 Novosibirsk, Russia
    }
\address[IKPUU]{Division of Nuclear and Particle Physics,
    Department of Physics and Astronomy,
    Uppsala University, Box 516, 75120 Uppsala, Sweden
    }
\address[IPJ]{Institute of Physics,
    Jagiellonian University, ul.\ Reymonta 4,
     30-059 Krak\'{o}w, Poland
     }
% ...
\address[ZELJ]{Zentralinstitut f\"ur Elektronik,
    Forschungszentrum J\"ulich,
    52425 J\"ulich, Germany
    }
% ...
\address[MS]{Institut f\"ur Kernphysik,
    Westf\"alische Wilhelms-Universit\"at M\"unster,
    Wilhelm--Klemm--Str.\ 9, 48149
    M\"unster, Germany
    }
\address[KEK]{High Energy Accelerator Research Organisation KEK,
    1--1 Oho, Tsukuba, Ibaraki 305-0801 Japan
    }
\address[Katow]{August Che{\l}kowski Institute of Physics,
    University of Silesia,
    Uniwersytecka 4, 40-007, Katowice, Poland
    }
% ...
\address[ASLodz]{Department of Cosmic Ray Physics,
    The Andrzej Soltan Institute for Nuclear Studies,
    ul.\ Uniwersytecka 5, 90-950 Lodz, Poland
    }
% ...
\address[NuJINR]{Dzhelepov Laboratory of Nuclear Problems,
    Joint Institute for Nuclear Physics,
    141980 Dubna, Russia
    }
% ...
\address[IMPCAS]{Institute of Modern Physics,
    Chinese Academy of Sciences,
    509 Nanchang Rd., 730000 Lanzhou, China
    }
\address[ASSwi]{Department of Physics Applications,
    The Andrzej Soltan Institute for Nuclear Studies,
    05-400 Otwock--\'{S}wierk, Poland
    }
%--------------------------------------------------
%\fntext[fnbb]{present address:}
\fntext[fndd]{deceased}
\fntext[fnmj]{present address: Laboratori Nazionali di Frascati,
    Via E.~Fermi 40, 00044 Frascati, Italy
    }
\fntext[fnik]{present address: Departement f\"ur Physik,
    Universit\"at Basel,
    Klingelbergstr.~82, 4056 Basel, Switzerland
    }
\fntext[fnvk]{present address:Institute for Nuclear Research,
    Prospect Nauki 47, 03680 Kyiv, Ukraine
    }
\fntext[fnnm]{present address: Technische Universit\"at Dortmund,
    Experimentelle Physik 5B,
    Otto-Hahn-Stra{\ss}e 4, 44227 Dortmund, Germany
    }
%\fntext[fnrn]{present address:}
\fntext[fnws]{present address: Unternehmensentwicklung und Au{\ss}enbeziehungen,
    Forschungszentrum J\"ulich,
    52425 J\"ulich, Germany
    }
%\fntext[fnku]{present address:}

\begin{abstract}
% Text of abstract
In the  first production run of  the WASA experiment at  COSY, the eta
decay  into  three  neutral   pions  was  measured  in  proton--proton
interactions at a  proton beam kinetic energy of  1.4~GeV.  The Dalitz
plot  of  the three  pions  was  studied  using 1.2$\times10^5$  fully
reconstructed events,  and the quadratic slope  parameter $\alpha$ was
determined  to be  $  -0.027  \pm 0.008({\rm  stat})  \pm 0.005  ({\rm
  syst})$.  The  result is  consistent with previous  measurements and
further  corroborates   the  importance  of   pion--pion  final  state
interactions.
\end{abstract}

\begin{keyword}
% keywords here, in the form: keyword \sep keyword
$\eta$ \sep $\eta \rightarrow 3 \pi^0$ \sep Dalitz plot
% PACS codes here, in the form: \PACS code \sep code
\PACS 13.25.-k \sep 13.75.Lb
% MSC codes here, in the form: \MSC code \sep code
% or \MSC[2008] code \sep code (2000 is the default)
\end{keyword}

\end{frontmatter}

% main text
\section{Introduction}
\label{section:introduction}

The  $\eta$ meson  plays a  special role  in understanding  low--energy
Quantum  Chromo Dynamics  (QCD). Chiral  symmetry,  its realization  in
hadron physics  at low  energies and the  role of  explicit chiral
symmetry breaking due to the  masses of the light quarks $(u,d,s)$ can
be investigated using $\eta$ decays.

The $\eta$ meson decays into three pions ($\eta\to\pi^0\pi^0\pi^0$ and
$\eta\to\pi^+\pi^-\pi^0$)  are  among   the  major  decay  modes  (the
branching    ratios    are    32.6\%    and    22.7\%,    respectively
\cite{Amsler:2008zz}) despite  the fact that isospin  is not conserved
in these  processes.  It  appears that these  decays are driven  by an
isospin breaking term  in the QCD Lagrangian which  is proportional to
the  quark  mass   difference  $m_d-m_u$  \cite{Bijnens:2002qy}.   Any
contribution    from   electromagnetic    processes    is   suppressed
\cite{Sutherland:1967vf,Baur:1995gc,Ditsche:2008ab}.    The  decay  is
closely related  to pion--pion scattering,  an  elementary low--energy
QCD process.  The lowest order  contribution to the decay mechanism is
given  by  Current  Algebra (CA)  \cite{Bardeen:1967ab,Osborn:1970nn}.
The predicted partial  decay width is more than  four times lower than
the measured value.  On the other hand, the distributions of the decay
products are described within a few percent accuracy.

\begin{figure}
\centerline{\includegraphics[width=0.7\linewidth]{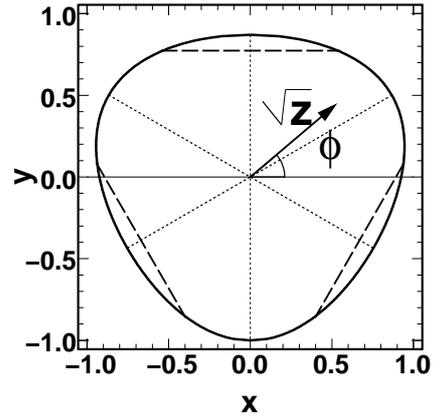}}
\caption{\label{fig:daldef}   Symmetrized    Dalitz   plot   for   the
  $\eta\to\pi^0\pi^0\pi^0$ decay. Dashed lines represent threshold for
  $\pi^0\pi^0\to\pi^-\pi^+$ rescattering. }
\end{figure}

The decay amplitude  of $\eta$ into three pions can be described
using the following two Dalitz variables:
\begin{equation}
\label{eqn:xydef}
  x\equiv\frac{1}{\sqrt{3}}\frac{T_1-T_2}{\langle T\rangle};\ \
y\equiv\frac{T_3}{\langle T\rangle}-1.
\end{equation}
Here $T_i$ are the kinetic energies  of the pions in the rest frame of
the $\eta$  meson, $3\langle T\rangle\equiv T_1+T_2+T_3=m_\eta-3m_\pi$
where  $m_\pi$  is  the  pion  mass  and  the  $\pi^0$,  $\pi^+$  mass
difference is neglected.  The boundaries  of the Dalitz plot are shown
in Fig.~\ref{fig:daldef}.

The decay  amplitude predicted by CA  is a linear function  of $y$ for
the   $\eta\to\pi^+\pi^-\pi^0$    decay   what   implies    that   the
$\eta\to\pi^0\pi^0\pi^0$ decay amplitude ($\bar{\cal A}$) is constant.
Experimentally, a  small deviation from a uniform  Dalitz Plot density
distribution is observed.  The lowest  term in the expansion about the
center of the Dalitz plot is given by:
\begin{equation}
 |\bar{\cal A}(z,\phi)|^2 = c(1 + 2 \alpha z)
\end{equation}
where $z$ and $\phi$ variables are related to $x$ and $y$ via:
\begin{equation}
  x=\sqrt{z}\cos\phi;\ \ y=\sqrt{z}\sin\phi.
\end{equation}
$c$ is  a constant  factor and  $\alpha$ is  the  quadratic slope
parameter.

The slope $\alpha$ was measured  to be negative and small, which leads
to  a decrease  of the  Dalitz plot  density at  the border  by  a few
percent.  The explanation of this  effect poses a challenge for Chiral
Perturbation Theory (ChPT),  the effective field theory of  QCD in the
low--energy region.   The leading order calculations  in ChPT coincide
with  CA  whereas   the  next--to--leading  order  (NLO)  calculations
\cite{Gasser:1984pr}  significantly  improve  the  agreement  for  the
partial decay width  but predict a small positive  value for $\alpha$.
The    large    uncertainty    of    the    recently    carried    out
next--to--next--to--leading       order       (NNLO)      calculations
\cite{Bijnens:2007pr} does not allow to  decide the sign of the slope.
The origin of  the non--uniform density distribution in  the $3 \pi^0$
Dalitz plot  are in part  pion--pion interactions in the  final state.
In  contrast  to  the   perturbative  calculations,  as  soon  as  the
rescattering   effects   are  considered   to   infinite  orders   via
unitarization  of  the  decay  amplitude  using  dispersion  relations
\cite{Kambor:1995yc,Beisert:2003zs} or iteration of the Bethe-Salpeter
equation \cite{Borasoy:2005du} the sign of $\alpha$ turns out to be in
accordance with experiment.   The calculations indeed predict negative
values for  $\alpha$ in agreement with the  experimental results which
are  summarized  in  Table  \ref{tab:alpha_table}  together  with  the
different theoretical predictions.

Furthermore, $\eta  \rightarrow 3 \pi$  decays are considered to  be a
source  of  precise  constraints  for  the  light  quark  mass  ratios
\cite{Leutwyler:1996qg}.   In a  recent approach  the  constraints are
derived entirely  from the experimental  partial decay width  and from
the    data   on    $\eta\to\pi^0\pi^0\pi^0$    Dalitz   plot    slope
\cite{Deandrea:2008px}.

The first precise experimental determination of the $\alpha$ parameter
was carried out by Crystal  Ball at AGS \cite{Tippens:2001fm} in 2001,
using a  pion beam  impinging on a  liquid hydrogen target.   In 2005,
KLOE released their  first result of a high  statistics measurement of
$\alpha$  (using  $e^{+}   e^{-}  \rightarrow  \Phi  \rightarrow  \eta
\gamma$)  \cite{Giovannella:2005rz} which  differed  from the  Crystal
Ball   result  by   three  standard   deviations.    The  CELSIUS/WASA
\cite{Bashkanov:2007iy} experiment has  confirmed the negative sign of
the slope parameter $\alpha$.   However, the achieved accuracy did not
allow  to resolve  the discrepancy  between the  Crystal Ball  and the
early KLOE results.  Very recently,  the KLOE data were reanalyzed and
a  new $\alpha$  value was  presented \cite{Ambrosino:2007wi}  that is
consistent  with Crystal  Ball.  However,  the final  results  are not
published yet.  The Crystal  Ball collaboration has recently collected
additional statistics  at the MAMI  facility that are  currently being
analyzed \cite{Starostin:2007ZZ}.

\begin{table*}[htb]
  \begin{center}
    \begin{tabular}{l ll r}
      \hline \hline \\
      Slope parameter \\ ~~$\alpha \pm {\rm stat} \pm {\rm syst}$ & Comment&year &  Reference    \\
\hline
      \\
      $-0.022\pm 0.023 $            &GAMS-2000&  (1984)      & \cite{Alde:1984wj}       \\%1984
      $-0.052\pm 0.017 \pm 0.010$   &Crystal Barrel& (1998) & \cite{Abele:1998yi}       \\%1998
      $-0.031\pm 0.004 \pm 0.004$   &Crystal Ball& (2001)    & \cite{Tippens:2001fm}    \\%2001
      $-0.014\pm 0.004 \pm 0.005$   &KLOE, preliminary& (2005)& \cite{Giovannella:2005rz} \\%2005
     $-0.027\pm 0.004 ~ ^{+0.004} _{-0.006}$   &KLOE, reanalysis& (2007) & \cite{Ambrosino:2007wi}  \\%2007
      $-0.026\pm 0.010 \pm 0.010$   &CELSIUS/WASA&  (2007)   & \cite{Bashkanov:2007iy}  \\%2005
      $-0.027 \pm  0.008 \pm 0.005$ & WASA-at-COSY& (2008) & this result\\
      \hline\\
      $ 0 $                 &CA&(1967)  & \cite{Osborn:1970nn}    \\
      $+0.015$              &ChPT NLO    &(1984)  & \cite{Gasser:1984pr}    \\
      $-(0.007\div0.014) $  &ChPT NLO+dispersive&(1995)  & \cite{Baur:1995gc}  \\
      $-0.007 $             &UChPT          &(2003)  & \cite{Beisert:2003zs}   \\
      $-0.031\pm 0.003 $    &UChPT fit      &(2005)  & \cite{Borasoy:2005du}   \\
      $+0.013\pm 0.032$             &ChPT NNLO      &(2007)  & \cite{Bijnens:2007pr}\\
      \hline \hline
    \end{tabular}
  \end{center}
  \caption[] {Overview of experimental  and theoretical results  for the slope parameter
    $\alpha$.
  \label{tab:alpha_table}}
\end{table*}

The phase space of the $\eta\to 3\pi^0$ decay covers the threshold for
the $\pi^0\pi^0\to\pi^+\pi^-$ rescattering process as a consequence of
the fact that the neutral pion mass is slightly lower than the charged
pion mass.   The boundaries  correspond to the  $\pi^0\pi^0$ invariant
mass being equal  to $2m_{\pi^{\pm}}$.  In 2004 a  cusp like structure
in  the $\pi^0\pi^0$  invariant mass  for  the $K^+\to\pi^0\pi^0\pi^+$
decay was  observed by the  NA48/2 collaboration \cite{Batley:2005ax}.
The  effect   was  predicted  already   in  1997  by   Mei{\ss}ner  et
al.  \cite{Meissner:1997fa}  and  the  interpretation  of  the  NA48/2
results  was  given  by  Cabibbo \cite{Cabibbo:2004gq}.   The  process
provides a  precise determination  of a combination  of the  $I=0$ and
$I=2$  pion--pion $s$-wave  scattering  lengths, $a_0$  and $a_2$.   A
similar   effect   was  observed   in   the   $K_L\to  3\pi^0$   decay
\cite{Abouzaid:2008js} and  it should be present also  in the $\eta\to
3\pi^0$  and  $\eta'\to\pi^0\pi^0\eta$  decays.  There  are  important
consequences  of this  phenomenon  for the  analysis  of the  $\eta\to
3\pi^0$                                                           decay
\cite{Cabibbo:2005ez,Gamiz:2006km,Colangelo:2006va,Bashkanov:2007iy,Bissegger:2007yq}.
The amplitude is no longer a function only of the $z$ variable but has
to explicitly depend on  the Mandelstam variables.  The description of
the amplitude as  a polynomial function will fail  in the cusp region.
For example  the dependence of  the $\phi$ averaged  amplitude squared
($|\bar{\cal A}|^2$):
\begin{equation}
|\bar{\cal    A}|^2\equiv    \frac{1}{2\pi}    \int_0^{2\pi}|\bar{\cal
  A}(z,\phi)|^2d\phi
\end{equation}
on the  $z$ variable  cannot be linear  for $0.6<z<0.9$.   However the
most  sensitive  variables to  search  for  the  cusp effect  are  the
invariant  masses  of the  pion  pairs  (or  kinetic energies  of  the
pions). They  correspond to the  projections onto the dotted  lines in
Fig.~\ref{fig:daldef}.

The results presented  in this paper are based  on a measurement using
proton--proton  interactions with the  WASA detection  system recently
installed  at the  Cooler  Synchrotron COSY  at the  Forschungszentrum
J\"ulich GmbH.

\section{The experiment}
\label{section:exp_setup}
The WASA  detection system \cite{Bargholtz:2008ze}  was transported to
COSY (COoler SYnchrotron) in  the Summer 2005 \cite{Adam:2004ch}.  The
COSY  facility at  J\"ulich offers  polarized and  phase  space cooled
proton   and   deuteron   beams   with   momenta   up   to   3.7~GeV/c
\cite{Prasuhn:1995ii,Maier:1997zj}.  The available beam momentum range
allows   the  production   of  $\pi$,   $K$,  $\eta$,   $\omega$,  and
$\eta\prime$   mesons  well   above  the   production   thresholds  in
proton-proton  and proton-deuteron  interactions.  Most  of  the final
states in  $pN$, $pd$, and $dd$  reactions can be detected  due to the
nearly $4\pi$ acceptance of the  WASA detector for charged and neutral
particles.

The WASA  detector \cite{Bargholtz:2008ze} was  designed and optimized
for  studies of  production and  decays  of light  mesons in  hadronic
interactions.   It consists  of  three main  components: \newline  the
Forward  Detector   -  used  for  tagging  and   triggering  on  meson
production,  the Central  Detector -  used for  measuring  neutral and
charged  meson decay products,  and the  unique pellet  target system.
The target beam consists of  30~$\mu$m diameter pellets of hydrogen or
deuterium, providing  a high target  density in the order  of $10^{15}
~atoms/cm^2$.

\begin{figure}
\centerline{\includegraphics[angle=-90,width=\linewidth]{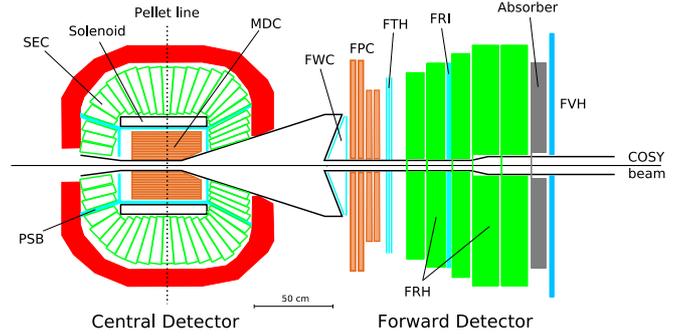}}
\caption{Schematic side view of the WASA detector setup at COSY.}
\end{figure}

The Central Detector surrounds  the interaction region and is designed
for detection  and identification  of photons, electrons,  and charged
pions.  It consists of an inner drift chamber (MDC), a superconducting
solenoid  providing the  magnetic field  for momentum  measurements, a
barrel of  thin plastic scintillators for  particle identification and
triggering,  and  an   electromagnetic  calorimeter.   The  amount  of
structural material is kept to  a minimum to reduce disturbance of the
particles.  The beryllium  beam pipe  wall is  1.2~$mm$ thick  and the
material of the superconducting solenoid corresponds to 0.18 radiation
lengths.

The  calorimeter (SEC)  is used  for detection  and  reconstruction of
particles  in the polar  angle range  from 20$^\circ$  to 169$^\circ$,
which is about 96$\%$  of the geometrical acceptance.  The calorimeter
consists of 1012 sodium  doped $C\!s\, I$ crystals.  The trapezoidally
shaped crystals with lengths from 20 to 30~$cm$ correspond to $\sim$16
radiation  lengths.   The  energy  resolution of  the  calorimeter  is
$\sigma{(E)}/E\approx         5\%/\sqrt{E         [{\rm        GeV}]}$
\cite{Bargholtz:2008ze}.

The forward detector is designed for detection of particles emitted in
the  polar  angles  from  3$^\circ$  to  18$^\circ$.   It  allows  for
identification and  reconstruction of protons from  the $pp\to pp\eta$
reaction close to threshold.   The precise track coordinates are given
by four  sets of straw proportional chambers  (FPC).  Kinetic energies
are reconstructed  using the $\Delta  E$ information in the  layers of
plastic scintillators of different thickness. In addition, the signals
are used  for triggering.   The kinetic energy  of the protons  can be
reconstructed  with a resolution  of $\sigma(T)/  T \sim  1.5-3\%$ for
kinetic energies below 400~MeV.

The  detector  setup  is nearly  the  same  as  used in  the  previous
CELSIUS/WASA experiment \cite{Bashkanov:2007iy}.  The main change is a
completely new readout  system with charge-to-digital converters based
on a  flash ADC \cite{Kleines:2006cy,Hejny:2007sv}.   In addition, the
fourth layer of the forward  range hodoscope (FRH) was replaced by two
new thicker layers (15~cm instead of 11~cm).

The $\eta$ mesons have been produced in proton--proton interactions at
a  beam proton  energy of  1.4~GeV (momentum  2.141~GeV/c).   The beam
energy corresponds to a center of mass excess energy of 56~MeV for the
$pp\to pp\eta$ reaction and the cross section is 10~$\mu$b.  The
results presented  here are  based on about  2.4 $pb^{-1}$  of data
collected during the first  WASA-at-COSY production run and correspond
to 94  hours of data  taking.  At the  trigger level, events  with two
tracks in  the forward  detector and  at least two  hit groups  in the
calorimeter with  energy deposits of  more than 50~MeV  were accepted.
In  addition,  a veto  on  signals in  the  plastic  barrel (PSB)  was
required, aiming to select only neutral particles.

\section{Data Analysis}
\begin{figure}[t]
      \centerline{\includegraphics[width=\linewidth]{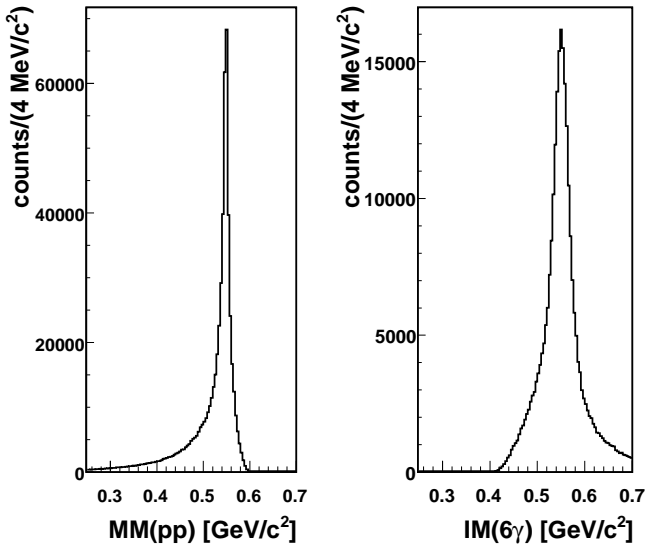}}
    \caption
    {Missing mass of two protons, $MM(pp)$ (left) and invariant mass of 
six photons, $IM(2\gamma)$ (right)     after selection of the 
$pp6\gamma$ final state.}
    \label{fig:immm}
\end{figure}

In the offline  analysis, the $pp 6\gamma$ final  state is selected by
requiring six hit clusters with  energy deposit at least 20~MeV in the
calorimeter  consistent with  photons  and two  proton  tracks in  the
forward detector.  The conditions provide a clean data sample of about
$8\cdot10^{5}$ events.  In order to select $\pi^0$ candidates from the
six reconstructed  photons, all  fifteen possible combinations  of the
three  photon  pairs  were  considered.   For  each  combination,  the
quantity $\chi^2_j$ is calculated:
\begin{equation}
  \chi^2_j \equiv \sum_{i}^{3} { {  ( IM(j,i)-m_{\pi^0} )^{2} } \over {
      \sigma^{2} } },\ \ \ \ \ \ j=1,2,3,...15
\end{equation}
where $IM(j,i)  $ is the  invariant mass of the  $i$-th $\gamma\gamma$
pair  for  the $j$-th  combination,  $\sigma  =  14$~MeV/c$^2$ is  the
resolution of the $\gamma\gamma$  invariant mass. The combination with
minimum  value  of  $\chi^2_j$   is  selected.   Only  events  with  $
\chi^2_j<15$ and a missing mass of two protons between 0.535~GeV/c$^2$
and 0.565~GeV/c$^2$ are kept.  Fig. \ref{fig:immm} shows the missing
mass of two protons (left panel) and the invariant mass of six photons
(right panel)  after applying  the cut on  the missing mass  and after
fitting the individual pions.

%
%------------- kinematic fit----------------------
%

\begin{figure}[h]
  \centerline{\includegraphics[width=\linewidth]{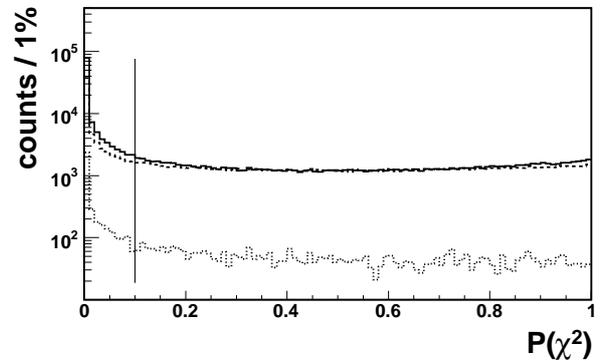}}
  \caption
  [probability  distribution] {The distribution  of the  kinematic fit
    probability  with   the  $\eta\to  3\pi^0$  decay  hypothesis
    ($P(\chi^2)$) for  the experimental data  (solid line), Monte
    Carlo  simulation  of the  signal  (dashed  line), and  background
    (dotted line).  The vertical line represents the selection for the
    final Dalitz plot analysis.  }
  \label{fig:kf_performance}
\end{figure}

In order to  reconstruct the $3\pi^0$ Dalitz plot,  a kinematic fit is
applied.   In  ref.   \cite{Bashkanov:2007iy},  the full  final  state
including  the two protons  was considered  in the  fitting procedure.
Here, only the $\eta\rightarrow 3 \pi^0$ decay system is fitted, based
on the reconstructed photon angles and momenta.  This approach reduces
systematic   uncertainty   due to proton  reconstruction.    Different
reconstruction uncertainties (as a  function of energy and angle) were
obtained  from a  full  GEANT Monte  Carlo  detector simulation,  with
resolution parameters matched to reproduce experimental distributions.
The  experimental  distribution   of  the  kinematic  fit  probability
($P(\chi^2)$)  is compared in  Fig.~\ref{fig:kf_performance} to
the Monte  Carlo simulation.  The simulation  includes background from
direct  three pion  production, $pp\to  pp\pi^0\pi^0\pi^0$,  the cross
section  was  obtained  by  interpolation  of  the  results  from  the
CELSIUS/WASA experiment  \cite{Pauly:2006pm}.  The relative  amount of
background  is  less  than  $4\%$  in the  final  event  sample  after
$P(\chi^2)>0.1$        cut        (vertical        line        in
Fig.~\ref{fig:kf_performance}).

\section{Extraction of  the slope parameter}

\begin{figure}[ht]
  \centerline{\includegraphics[width=\linewidth]{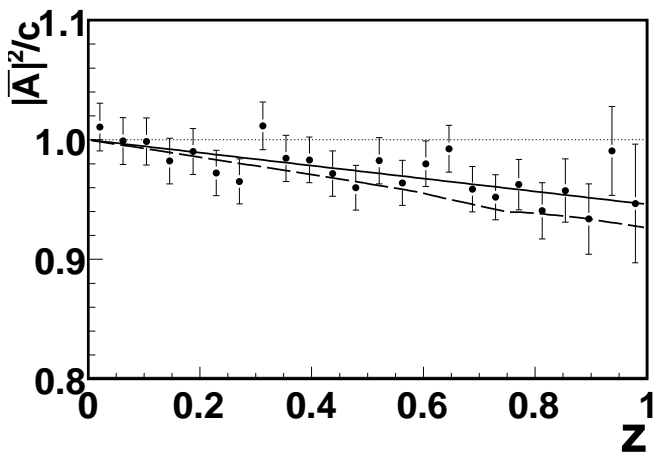}}
  \caption{Extracted dependence of  $|\bar{\cal A}|^2$ on  the  $z$ variable.
  The  solid line  is a $c(1+2\alpha z)$ fit.  The dashed line is a prediction
  of the cusp effect \cite{Rusetsky:2008ab}.
  \label{fig:dp_dd}}
\end{figure}

Based  on  the analysis  procedure,  the  efficiency corrected  radial
density    distribution    $|\bar{\cal     A}|^2$    is    shown    in
Fig.~\ref{fig:dp_dd}.   The  efficiency  correction is  obtained  by
dividing  the measured  $z$ distribution  by the  result of  the Monte
Carlo  simulation, assuming  $\alpha=0$.   A linear  fit  to the  data
points is applied to extract  the slope parameter $\alpha$, yielding a
statistical  uncertainty of  $\sigma(\alpha) =  8\cdot  10^{-3}$.  The
achieved resolution in z is $\sigma(z)$=0.055.

The systematic uncertainty  of the result is estimated  by varying one
by   one  all  parameters   that  are   important  in   the  analysis.
Table~\ref{tab:sys_sum}  summarizes   some  of  these   studies.   The
dominant  contribution  comes  from   an  uncertain  fraction  of  the
interactions with gas stemming from pellet evaporation or with pellets
bouncing  in  the beam  pipe.   This causes  a  spread  in the  vertex
position.\newline The combinatoric purity of the selected event sample
depends on the cut on $\chi^{2}_j$ probability of the best combination
and  on  the  confidence level  of  the  fit.   A variation  of  these
parameters   has  only   a  minor   effect  on   the   obtained  slope
result.\newline Another source of  systematic errors is imposed by the
remaining background.   This effect is  under control by  changing the
width of  the proton--proton missing  mass cut and,  hence, varying the
relative amount  of background in the range  from 4 \% to  12 \%.  The
observed  effect  contributes  with   0.002  to  the  slope  parameter
uncertainty.
\newline
The kinematic  fit performance relies on the  precise understanding of
the  errors of reconstructed  quantities. This  effect was  studied by
replacing the polar  angle dependent parametrization of reconstruction
uncertainties with a much simpler polar angle independent description,
again  showing  only   a  small  effect  on  the   result  (see  table
\ref{tab:sys_sum}).

\begin{table}[h]
  \begin{center}
    \begin{tabular}{ll}
        \\ \hline\hline \\
        Source of systematic uncertainty  & RMS \\
        \hline
        Combinatoric background           & $ 0.001 $ \\
        Missing mass cut                  & $ 0.002 $ \\
        Background 3$\pi^0$ production    & $ 0.001 $ \\
        Vertex position                   & $ 0.004 $ \\
        Confidence level cut              & $ 0.001 $ \\
        Error parametrization for the kinematical fit & $ 0.002 $ \\
        \\ \hline
        Overall systematic uncertainty & $0.005$
        \\ \hline\hline \\
    \end{tabular}
  \end{center}
  \caption{The  main contributions  to the  systematic  uncertainty as
    discussed  in the  text.  The  overall systematic  uncertainty was
    calculated  as  the square  root  of  the  summed squares  of  all
    contributions.\label{tab:sys_sum}  }
\end{table}

Taking into account the systematic studies, the final result for the
extracted slope parameter $\alpha$ is
\begin{center}
  $\alpha = -0.027 \pm 0.008(stat) \pm 0.005 (syst)$
\end{center}
based  on  120000  events.   The  result agrees  within  one  standard
deviation with the high statistics measurements of the slope parameter
performed  by the KLOE  collaboration \cite{Ambrosino:2007wi}  and the
Crystal   Ball  collaboration   \cite{Tippens:2001fm}.   It   is  also
consistent with the previous measurement performed by the CELSIUS/WASA
collaboration   \cite{Bashkanov:2007iy}.   The   shape   of  the   $z$
distribution  including   the  cusp   that  emerges  from   a  virtual
$\pi^+\pi^-$   intermediate  state   followed   by  a   $\pi^+\pi^-\to
\pi^0\pi^0$    transition,     can    be    calculated     from    the
$\eta\to\pi^+\pi-\pi^0$   amplitude  and   the   $\pi-\pi$  scattering
lengths. The cusp effect leads to  a broad local minimum in the radial
density of the Dalitz plot  for $0.6<z<0.9$ due to the contribution of
the  regions where  the  invariant mass  of  the two  pions less  than
$2m_{\pi^\pm}$  (see  Fig.~\ref{fig:daldef}).   The dashed  line  in
Fig.~\ref{fig:dp_dd}  was obtained using  a parameterization  of the
$\eta\to\pi^+\pi^-\pi^0$ decay  amplitude from the  KLOE collaboration
\cite{Ambrosino:2008ht} and  a nonrelativistic effective  field theory
\cite{Bissegger:2007yq,Rusetsky:2008ab}.    It   is   seen  that   the
statistical precision  of the data is insufficient  to investigate the
cusp effect.

\section{Outlook}
One motivation  for the presented  study was the  striking discrepancy
between  KLOE and  Crystal Ball  results for  the $\eta  \rightarrow 3
\pi^0$ Dalitz  plot slope parameter  $\alpha$.  Now, after  the recent
reanalysis of the  KLOE data, the three major  experiments focusing on
the measurement of $\eta$ decays,  KLOE at DAPHNE in Frascati, Crystal
Ball now  installed at  MAMI in Mainz,  and WASA-at-COSY  in J\"ulich,
obtain consistent results.  The non--zero value of the slope parameter
is  clear indication  for  the importance  of  final state  pion--pion
interactions.   The theoretical  understanding of these processes has
advanced  significantly,  also  triggered  by  the  firm  experimental
situation.  The Dalitz plot  slope parameter provides a very sensitive
test of the ChPT predictions.  The ChPT NLO and NNLO calculations that
assume $m_{\pi^0}=m_{\pi^\pm}$ in the  loops, indicate a positive sign
of  the slope  parameter.  The  uncertainty is  however large  and the
negative sign is not excluded \cite{Bijnens:2007pr}.

Among the  experiments studying $\eta$  decays, WASA is  presently the
only experiment focusing on hadronic  production of eta mesons in $pp$
or $pd$ scattering.  The experiment combines the high production cross
section especially in the $pp$  reaction with the capability to run at
high luminosities  up to $10^{32} cm^{-2}s^{-1}$.  This  will allow to
collect high  statistics data samples of $\eta$  mesons ($10^7$ decays
and  more) with  WASA-at-COSY.  High  statistics data  samples provide
further  increased accuracy  and  sensitivity in  the  study of  small
effects,  like  the cusp  structure  in  $\eta  \rightarrow 3  \pi^0$.
Recently   WASA-at-COSY  has   collected  a   large  data   sample  of
$pd\to{\rm{^3He}}\eta$  events  with low  background  and an  unbiased
trigger    requirement,    enabling    studies    {\em    e.g.}     of
$\eta\to\pi^0\pi^0\pi^0$     and    $\eta\to\pi^+\pi^-\pi^0$    decays
simultaneously.

\section{Acknowledgments}

This  work was in  part supported  by: the  Forschungszentrum J\"ulich
including the  COSY-FFE program, the European Community  under the FP6
program  (Hadron Physics, RII3-CT-2004-506078),  the German  BMBF, the
german-indian  DAAD-DST exchange  program, VIQCD  (VH-VI-231)  and the
German Research Foundation (DFG).

We gratefully acknowledge the financial  support given by the Knut and
Alice Wallenberg Foundation, the Swedish Research Council, the G\"oran
Gustafsson  Foundation,  the Polish  Ministry  of  Science and  Higher
Education    under    grants    PBS   7P-P6-2/07,    3240/H03/2006/31,
1202/DFG/2007/03,

We also  want to thank the  technical and administration  staff at the
Forschungszentrum J\"ulich and at the participating institutes.

This work is part of the PhD Thesis of P. Vlasov.

%
% The Appendices part is started with the command \appendix;
% appendix sections are then done as normal sections
% \appendix

%\newpage
%\pagestyle{empty}

%\bibliographystyle{is-unsrt}
%\bibliographystyle{h-elsevier3}
\bibliographystyle{utphys}
\bibliography{papers}{}

\providecommand{\href}[2]{#2}\begingroup\raggedright\begin{thebibliography}{10}

\bibitem{Amsler:2008zz}
{\bf Particle Data Group} Collaboration, C.~Amsler {\em et al.}
\href{http://dx.doi.org/10.1016/j.physletb.2008.07.018}{{\em Phys. Lett.} {\bf
  B667} (2008)  1}.
%%CITATION = PHLTA,B667,1;%%.

\bibitem{Bijnens:2002qy}
J.~Bijnens and J.~Gasser
  \href{http://dx.doi.org/10.1238/Physica.Topical.099a00034}{{\em Phys.
  Scripta} {\bf T99} (2002)  34},
\href{http://arxiv.org/abs/hep-ph/0202242}{{\tt arXiv:hep-ph/0202242}}.
%%CITATION = HEP-PH/0202242;%%.

\bibitem{Sutherland:1967vf}
D.~G. Sutherland
\href{http://dx.doi.org/10.1016/0550-3213(67)90180-0}{{\em Nucl. Phys.} {\bf
  B2} (1967)  433}.
%%CITATION = NUPHA,B2,433;%%.

\bibitem{Baur:1995gc}
R.~Baur, J.~Kambor, and D.~Wyler
  \href{http://dx.doi.org/10.1016/0550-3213(95)00643-5}{{\em Nucl. Phys.} {\bf
  B460} (1996)  127},
\href{http://arxiv.org/abs/hep-ph/9510396}{{\tt arXiv:hep-ph/9510396}}.
%%CITATION = HEP-PH/9510396;%%.

\bibitem{Ditsche:2008ab}
C.~Ditsche, B.~Kubis, and U.-G. Mei{\ss}ner 2008.
\newblock in preparation.

\bibitem{Bardeen:1967ab}
W.~A. Bardeen, L.~S. Brown, B.~W. Lee, and H.~T. Nieh {\em Phys. Rev. Lett.}
  {\bf 18} (1967) no.~25, 1170.

\bibitem{Osborn:1970nn}
H.~Osborn and D.~J. Wallace
\href{http://dx.doi.org/10.1016/0550-3213(70)90194-X}{{\em Nucl. Phys.} {\bf
  B20} (1970)  23}.
%%CITATION = NUPHA,B20,23;%%.

\bibitem{Gasser:1984pr}
J.~Gasser and H.~Leutwyler
\href{http://dx.doi.org/10.1016/0550-3213(85)90494-8}{{\em Nucl. Phys.} {\bf
  B250} (1985)  539}.
%%CITATION = NUPHA,B250,539;%%.

\bibitem{Bijnens:2007pr}
J.~Bijnens and K.~Ghorbani
  \href{http://dx.doi.org/10.1088/1126-6708/2007/11/030}{{\em JHEP} {\bf 11}
  (2007)  030},
\href{http://arxiv.org/abs/0709.0230}{{\tt arXiv:0709.0230 [hep-ph]}}.
%%CITATION = 0709.0230;%%.

\bibitem{Kambor:1995yc}
J.~Kambor, C.~Wiesendanger, and D.~Wyler
  \href{http://dx.doi.org/10.1016/0550-3213(95)00676-1}{{\em Nucl. Phys.} {\bf
  B465} (1996)  215},
\href{http://arxiv.org/abs/hep-ph/9509374}{{\tt arXiv:hep-ph/9509374}}.
%%CITATION = HEP-PH/9509374;%%.

\bibitem{Beisert:2003zs}
N.~Beisert and B.~Borasoy
  \href{http://dx.doi.org/10.1016/S0375-9474(02)01585-3}{{\em Nucl. Phys.} {\bf
  A716} (2003)  186},
\href{http://arxiv.org/abs/hep-ph/0301058}{{\tt arXiv:hep-ph/0301058}}.
%%CITATION = HEP-PH/0301058;%%.

\bibitem{Borasoy:2005du}
B.~Borasoy and R.~Ni{\ss}ler {\em Eur. Phys. J.} {\bf A26} (2005)  383,
\href{http://arxiv.org/abs/hep-ph/0510384}{{\tt arXiv:hep-ph/0510384}}.
%%CITATION = HEP-PH/0510384;%%.

\bibitem{Leutwyler:1996qg}
H.~Leutwyler \href{http://dx.doi.org/10.1016/0370-2693(96)00386-3}{{\em Phys.
  Lett.} {\bf B378} (1996)  313},
\href{http://arxiv.org/abs/hep-ph/9602366}{{\tt arXiv:hep-ph/9602366}}.
%%CITATION = HEP-PH/9602366;%%.

\bibitem{Deandrea:2008px}
A.~Deandrea, A.~Nehme, and P.~Talavera
  \href{http://dx.doi.org/10.1103/PhysRevD.78.034032}{{\em Phys. Rev.} {\bf
  D78} (2008)  034032},
\href{http://arxiv.org/abs/0803.2956}{{\tt arXiv:0803.2956 [hep-ph]}}.
%%CITATION = 0803.2956;%%.

\bibitem{Tippens:2001fm}
{\bf Crystal Ball} Collaboration, W.~B. Tippens {\em et al.}
\href{http://dx.doi.org/10.1103/PhysRevLett.87.192001}{{\em Phys. Rev. Lett.}
  {\bf 87} (2001)  192001}.
%%CITATION = PRLTA,87,192001;%%.

\bibitem{Giovannella:2005rz}
{\bf KLOE} Collaboration, S.~Giovannella {\em et al.} in {\em Proceedings of La
  Thuile 2005, Results and perspectives in particle physics}, p.~241.
\newblock Rencontres de Moriond, 2005.
\newblock
\href{http://arxiv.org/abs/hep-ex/0505074}{{\tt arXiv:hep-ex/0505074}}.
\newblock
%%CITATION = HEP-EX/0505074;%%.

\bibitem{Bashkanov:2007iy}
{\bf CELSIUS/WASA} Collaboration, M.~Bashkanov {\em et al.}
  \href{http://dx.doi.org/10.1103/PhysRevC.76.048201}{{\em Phys. Rev.} {\bf
  C76} (2007)  048201},
\href{http://arxiv.org/abs/0708.2014}{{\tt arXiv:0708.2014 [nucl-ex]}}.
%%CITATION = 0708.2014;%%.

\bibitem{Ambrosino:2007wi}
{\bf KLOE} Collaboration, F.~Ambrosino {\em et al.} in {\em Proceedings of LP07
  conference}, pp.~S8--356.
\newblock Kyungpook National University Press, 2007.
\newblock
\href{http://arxiv.org/abs/0707.4137}{{\tt arXiv:0707.4137 [hep-ex]}}.
\newblock
%%CITATION = 0707.4137;%%.

\bibitem{Starostin:2007ZZ}
A.~Starostin and S.~Prakhov.
In the Proceedings of 11th International Conference on Meson-Nucleon Physics
  and the Structure of the Nucleon (MENU 2007), Julich, Germany, 10-14 Sep
  2007.
%%CITATION = ECONF,C070910,114;%%.

\bibitem{Alde:1984wj}
{\bf Serpukhov-Brussels-Annecy(LAPP)} Collaboration, D.~Alde {\em et al.}
\href{http://dx.doi.org/10.1007/BF01547921}{{\em Z. Phys.} {\bf C25} (1984)
  225}.
%%CITATION = ZEPYA,C25,225;%%.

\bibitem{Abele:1998yi}
{\bf Crystal Barrel} Collaboration, A.~Abele {\em et al.}
\href{http://dx.doi.org/10.1016/S0370-2693(97)01377-4}{{\em Phys. Lett.} {\bf
  B417} (1998)  193}.
%%CITATION = PHLTA,B417,193;%%.

\bibitem{Batley:2005ax}
{\bf NA48/2} Collaboration, J.~R. Batley {\em et al.} {\em Phys. Lett.} {\bf
  B633} (2006)  173,
\href{http://arxiv.org/abs/hep-ex/0511056}{{\tt hep-ex/0511056}}.
%%CITATION = HEP-EX/0511056;%%.

\bibitem{Meissner:1997fa}
U.-G. Mei{\ss}ner, G.~M{\"u}ller, and S.~Steininger
  \href{http://dx.doi.org/10.1016/S0370-2693(97)00666-7}{{\em Phys. Lett.} {\bf
  B406} (1997)  154}, \href{http://arxiv.org/abs/hep-ph/9704377}{{\tt
  arXiv:hep-ph/9704377}}.
Erratum-ibid.B407:454,1997.
%%CITATION = HEP-PH/9704377;%%.

\bibitem{Cabibbo:2004gq}
N.~Cabibbo {\em Phys. Rev. Lett.} {\bf 93} (2004)  121801,
\href{http://arxiv.org/abs/hep-ph/0405001}{{\tt hep-ph/0405001}}.
%%CITATION = HEP-PH/0405001;%%.

\bibitem{Abouzaid:2008js}
{\bf KTeV} Collaboration, E.~Abouzaid {\em et al.}
  \href{http://dx.doi.org/10.1103/PhysRevD.78.032009}{{\em Phys. Rev.} {\bf
  D78} (2008)  032009},
\href{http://arxiv.org/abs/0806.3535}{{\tt arXiv:0806.3535 [hep-ex]}}.
%%CITATION = 0806.3535;%%.

\bibitem{Cabibbo:2005ez}
N.~Cabibbo and G.~Isidori
  \href{http://dx.doi.org/10.1088/1126-6708/2005/03/021}{{\em JHEP} {\bf 03}
  (2005)  021},
\href{http://arxiv.org/abs/hep-ph/0502130}{{\tt arXiv:hep-ph/0502130}}.
%%CITATION = HEP-PH/0502130;%%.

\bibitem{Gamiz:2006km}
E.~Gamiz, J.~Prades, and I.~Scimemi
  \href{http://dx.doi.org/10.1140/epjc/s10052-006-0201-7}{{\em Eur. Phys. J.}
  {\bf C50} (2007)  405},
\href{http://arxiv.org/abs/hep-ph/0602023}{{\tt arXiv:hep-ph/0602023}}.
%%CITATION = HEP-PH/0602023;%%.

\bibitem{Colangelo:2006va}
G.~Colangelo, J.~Gasser, B.~Kubis, and A.~Rusetsky
  \href{http://dx.doi.org/10.1016/j.physletb.2006.05.017}{{\em Phys. Lett.}
  {\bf B638} (2006)  187},
\href{http://arxiv.org/abs/hep-ph/0604084}{{\tt arXiv:hep-ph/0604084}}.
%%CITATION = HEP-PH/0604084;%%.

\bibitem{Bissegger:2007yq}
M.~Bissegger, A.~Fuhrer, J.~Gasser, B.~Kubis, and A.~Rusetsky
  \href{http://dx.doi.org/10.1016/j.physletb.2007.11.008}{{\em Phys. Lett.}
  {\bf B659} (2008)  576},
\href{http://arxiv.org/abs/0710.4456}{{\tt arXiv:0710.4456 [hep-ph]}}.
%%CITATION = 0710.4456;%%.

\bibitem{Bargholtz:2008ze}
{\bf CELSIUS/WASA} Collaboration, C.~Bargholtz {\em et al.}
  \href{http://dx.doi.org/10.1016/j.nima.2008.06.011}{{\em Nucl. Instrum.
  Meth.} {\bf A594} (2008)  339},
\href{http://arxiv.org/abs/0803.2657}{{\tt arXiv:0803.2657 [nucl-ex]}}.
%%CITATION = 0803.2657;%%.

\bibitem{Adam:2004ch}
{\bf WASA-at-COSY} Collaboration, H.~H. Adam {\em et al.}
\href{http://arxiv.org/abs/nucl-ex/0411038}{{\tt arXiv:nucl-ex/0411038}}.
%%CITATION = NUCL-EX/0411038;%%.

\bibitem{Prasuhn:1995ii}
D.~Prasuhn {\em et al.}
\href{http://dx.doi.org/10.1016/0168-9002(95)00309-6}{{\em Nucl. Instrum.
  Meth.} {\bf A362} (1995)  16}.
%%CITATION = NUIMA,A362,16;%%.

\bibitem{Maier:1997zj}
R.~Maier
\href{http://dx.doi.org/10.1016/S0168-9002(97)00324-0}{{\em Nucl. Instrum.
  Meth.} {\bf A390} (1997)  1}.
%%CITATION = NUIMA,A390,1;%%.

\bibitem{Kleines:2006cy}
H.~Kleines {\em et al.}
\href{http://dx.doi.org/10.1109/TNS.2006.873305}{{\em IEEE Trans. Nucl. Sci.}
  {\bf 53} (2006)  893}.
%%CITATION = IETNA,53,893;%%.

\bibitem{Hejny:2007sv}
H.~Kleines {\em et al.}
\href{http://dx.doi.org/10.1109/TNS.2007.914033}{{\em IEEE Trans. Nucl. Sci.}
  {\bf 55} (2008)  261}.
%%CITATION = IETNA,55,261;%%.

\bibitem{Pauly:2006pm}
{\bf CELSIUS/WASA} Collaboration, C.~Pauly {\em et al.}
  \href{http://dx.doi.org/10.1016/j.physletb.2007.04.004}{{\em Phys. Lett.}
  {\bf B649} (2007)  122},
\href{http://arxiv.org/abs/nucl-ex/0602006}{{\tt arXiv:nucl-ex/0602006}}.
%%CITATION = NUCL-EX/0602006;%%.

\bibitem{Rusetsky:2008ab}
A.~Rusetsky, A.~Kupsc, and C.-O. Gullstrom 2008.
\newblock in preparation.

\bibitem{Ambrosino:2008ht}
{\bf KLOE} Collaboration, A.~Antonelli {\em et al.}
  \href{http://dx.doi.org/10.1088/1126-6708/2008/05/006}{{\em JHEP} {\bf 05}
  (2008)  006},
\href{http://arxiv.org/abs/0801.2642}{{\tt arXiv:0801.2642 [hep-ex]}}.
%%CITATION = 0801.2642;%%.

\end{thebibliography}\endgroup

% \begin{thebibliography}{00}       %original
% \bibitem{label}           %original
%   Text of bibliographic item      %original
% \bibitem{}                %original
% \end{thebibliography}         %original

\end{document}